\documentclass{article}  
\usepackage{spadre2008}
\usepackage[pdftex]{graphicx}
\frompage{000} \topage{000}                                              
\def\rightmark{Viscosity at RHIC: Theory and Practice}
\def\leftmark{S. Pratt}

\title{Viscosity at RHIC: Theory and Practice} 
\authors{
{Scott Pratt %
}\\[2.812mm]
{\normalsize
\hspace*{-8pt}Department of Physics \&Astronomy, Michigan State University,
East Lansing, MI 48824 ~~USA\\[0.2ex] 
}}
 
\abstract{Hydrodynamic behavior and the associated discussions of viscosity at RHIC has inspired a renaissance in 
modeling viscous hydrodynamics. An explanation of Israel-Stewart hydrodynamics is presented here, with an emphasis on 
the tangible benefits compared to Navier Stokes. }

 
\begin{document}
 
\maketitle
\setcounter{page}{1}

\section{Introduction}\label{intro}

Strong collective flow observed at RHIC has validated the use of hydrodynamics to model the collision for much of the reaction. However, even with the small inferred viscosities, viscous effects are of the order of 10-20 percent. In the last year several significant advances have been made regarding the formalism, implementation and understanding of viscous relativistic hydrodynamics for RHIC. A consensus has been building that Israel-Stewart treatments are superior to the traditional Navier-Stokes approach, both for numerical and physical reasons. In this talk, I will explain and motivate the Israel-Stewart approach, then review the observable manifestations of viscosity in high-energy heavy-ion collisions with an emphasis on explaining femtoscopic measurements of source size. 

Viscous effects refer to the alteration of the stress-energy tensor due to the system's inability to maintain equilibrium. For equilibrated systems, the spatial components of the stress-energy tensor in the fluid's rest frame are:
\begin{equation}
T_{ij}=P(\epsilon,\vec{\rho})\delta_{ij}.
\end{equation}
Here, the pressure is a function of the energy density and the conserved charges $\vec{\rho}$. For rapidly expanding systems equilibrium cannot be maintained and $T^{ij}$ differs for a variety of physical reasons, \cite{Pratt:2006ss,Paech:2006st}:
\begin{enumerate}
\item Local kinetic distributions become anisotropic due to anisotropic flows. This is the text book case which leads to the shear viscosity being proportional to the mean free path \cite{weinberg}.
\item Interactions and correlations that extend over a finite range contribute to both the shear and bulk viscosities \cite{Cheng:2001dz}. For interactions of a finite range $\ell$, matter at ${\bf r}=0$ will be heated by the collective flow of the matter at distance $\ell$.
\item Near $T_c$, mean fields may find it difficult to adjust to rapidly changing equilibrium values. This results in peaks to the bulk viscosity near $T_c$ \cite{Paech:2006st,Karsch:2007jc}.
\item Chemical populations, especially for massive particles, fall out of equilibrium for rapidly expanding systems, resulting in fugacities. These effects can be incorporated either by explicity treating the unequilibrated populations as dynamical charges, or as a bulk viscosity.
\item The system may not have relaxed from initial conditions for $T^{ij}$. Most notably, at early times the stress-energy tensor might be dominated by longitudinal color fields \cite{Krasnitz:2002mn,Cheng:2001dz}, which are characterized by large transverse components of $T^{ij}$ and a small, perhaps even negative, longitudinal component.
\end{enumerate}

Historically, viscous effects have been most often incorporated through the Navier-Stokes equation,
\begin{equation}
T^{ij}=P\delta_{ij}-\zeta \nabla\cdot v-\eta\left(\partial_iv_j+\partial_jv_i-(2/3)\nabla\cdot v\right).
\end{equation}
Like the pressure, the bulk and shear viscosities, $\zeta$ and $\eta$, are functions of $\epsilon$ and $\vec{\rho}$. The coefficients represent fundamental properties of the system that can be related to correlations in an equilibrated system through Kubo relations,
\begin{eqnarray}
\eta&=&\frac{1}{i\hbar}\int d^3r \int_0^\infty dt~t\langle \left[T^{xy}({\bf r}'=0,t'=0),T^{xy}({\bf r},t)\right]\rangle,\\
\nonumber
\zeta&=&\frac{1}{i\hbar}\int d^3r \int_0^\infty dt~t\langle \left[\delta \bar{T}({\bf r}'=0,t'=0)\delta\bar{T}({\bf r},t)\right]\rangle,\\
\nonumber
\delta\bar{T}&\equiv&\frac{1}{3}\left(T_{xx}+T_{yy}+T_{zz}\right)-P-c_s^2\left(T_{00}-\langle T_{00}\rangle\right).
\end{eqnarray}
Lattice Gauge theory provides such correlations of the stress energy tensor, but only for imaginary times, $0<t<i\beta$. The difficulties of extracting to real times is discussed in \cite{Pratt:2007gj}.

Navier-Stokes treatments can be problematic for two reasons. First, the treatments can have super-luminar or unstable behavior for short wavelength modes. Secondly, the treatments implicitly assume that the anisotropy is determined by the instantaneous velocity gradient, rather than the past history or initial conditions. This restricts the application to small velocity gradients and to times sufficiently long so that memory of the initial state of the stress-energy tensor is lost. Israel Stewart approaches \cite{israelstewart,muronga,koide,baierromatschke,heinzchaudhuri,songheinz,romatschke,Pratt:2007gj} avoid both shortcomings.

Israel-Stewart hydrodynamics is predicated on the assumption that the non-equilibrium part of $T_{ij}$ will exponentially relax toward Navier-Stokes values,
\begin{equation}
\frac{D}{Dt}a_i=\frac{-1}{\tau_a}\left(a_i-a_i^{\rm (NS)}\right),~~~~~
\frac{D}{Dt}b=\frac{-1}{\tau_b}\left(b-b^{\rm (NS)}\right),
\end{equation}
where $a_i$ and $b$ are the non-equilibrium parts of the stress energy tensor,
\begin{eqnarray}
\label{eq:abdef}
b&\equiv&\frac{1}{3\sigma_b}\left(T_{xx}+T_{yy}+T_{zz}\right)-P,\\
\nonumber
a_1&\equiv&\frac{1}{2\sigma_a}\left(T_{xx}-T_{yy}\right),~~~
a_2\equiv\frac{1}{\sigma_a\sqrt{12}}\left(T_{xx}+T_{yy}-2T_{zz}\right),\\
\nonumber
a_3&\equiv&\frac{T_{xy}}{\sigma_a},~~a_4\equiv\frac{T_{xz}}{\sigma_a},~~a_5\equiv\frac{T_{yz}}{\sigma_a}.
\end{eqnarray}
Here, the exponential scaling functions are not arbitrary, but like the viscosity can be expressed as correlations,
\begin{equation}
\label{eq:abconstraints}
\sigma_b^2=\int d^3r\langle \delta \bar{T}({\bf r}'=0)\delta \bar{T}({\bf r})\rangle,~~~
\sigma_a^2=\int d^3r\langle T_{xy}({\bf r}'=0)T_{xy}({\bf r})\rangle.
\end{equation}
Furthermore, the relaxation times are related to the fluctuations and to the viscosities,
\begin{equation}
\label{eq:tauconstraints}
\tau_b=\zeta T/\sigma_b^2,~~~\tau_a=\eta T/\sigma_a^2.
\end{equation}
The constraints in Eq.s (\ref{eq:abconstraints}) and Eq. (\ref{eq:tauconstraints}) follow from the constraints that entropy production must always be positive, or in the case of Eq. (\ref{eq:tauconstraints}), from linear response theory \cite{Pratt:2007gj,jou}. Unlike the viscous coefficients, $\sigma_a^2$ and $\sigma_b^2$ are equal-time correlations and can be calculated from lattice gauge theory without analytic continuation. Thus, relaxation times can be determined from lattice with the same confidence as viscosity coefficients.

Numerical problems with Navier-Stokes hydrodynamics derive from the parabolic nature of the equations, which like the diffusion equation has arbitrarily fast modes \cite{israelstewart,muronga}. The odd behavior occurs for high wave numbers and disappears for relaxation times of order, or larger, than a characteristic time $\tau_{\rm char}\equiv \frac{\eta}{hc_s^2}$, where $h$ is the enthalpy density and $c_s$ is the speed of sound \cite{Pratt:2007gj}. If one uses Eq. \ref{eq:tauconstraints} to estimate the relaxation times in terms of the fluctuations of the stress energy tensor, and if one then calculates the fluctuation and the enthalpy for independent particles, one can calculate both the characteristic time and the relaxation times, which both are proportional to $\eta$ for shear waves. The ratio of the two times is independent of $\eta$ and comfortably above the values where singular behavior might occur. Thus, if one follows the prescriptions for generating relaxation times consistent with the viscosity and fluctuations, the relaxation times should be sufficiently long to avoid numerical instabilities.

Elliptic flow is the most common observable to be associated with viscosity. However, viscosity also significantly affect spectra and correlations \cite{Paech:2006st,romatschke}. These manifestations can be understood by considering transverse acceleration at early times, which is driven by the transverse components of the stress energy tensor, $T_{xx}$ and $T_{yy}$. Since the velocity gradient is largely longitudinal, the effect of shear is to lower the longitudinal component, $T_{zz}$, while increasing the transverse components. From a variety of perspectives, it appears that an early boost of transverse collective flow is needed to explain the $R_{\rm out}/R_{\rm side}$ ratio measured in correlations analyses \cite{romatschke,Gyulassy:2007zz,Li:2007yd}. Even though hydrodynamics, even viscous hydrodynamics, is inapplicable for the earliest times, the equation for transverse acceleration, $(\epsilon+T_{xx})Dv/Dt=-\partial T_{xx}$, is always valid. In fact, for non-interacting longitudinal classical fields, $T_{xx}=T_{yy}=\epsilon$, three times higher than for an ultra-relativistic gas. A non-equilibrium system dominated by longitudinal classical fields is the most effective means to jump start transverse flow.

In the context of elliptic flow, the behavior of non-interacting fields contrasts with that of non-interacting particles. For non-interacting particles, strong transverse collective flow also develops, but is accompanied by a canceling anisotropy in the stress-energy tensor as measured in the frame of the flowing fluid. I.e., if one were looking in the $x$ direction, $T_{xx}$ would be less than $T_{yy}$ if measured in the frame of the fluid. However for classical longitudinal fields, the stress energy tensor remains isotropic in the rest frame, allowing significant elliptic flow to be generated within the first 1 fm/$c$. Thus, opposite much of the current wisdom, delayed thermalization can amplify both elliptic and radial flow.

Bulk viscosity is only expected to be significant near $T_c$ \cite{Paech:2006st,Karsch:2007jc}. This has the effect of increasing entropy and slowing the expansion, especially at the surface of the fireball. Friction diminishes collective flow, thus modestly reducing the mean transverse momentum. The increase in the entropy is then accommodated through increased source volumes as determined by two-particle correlations.

\begin{figure}[htb]
\centerline{\includegraphics[width=0.5\textwidth]{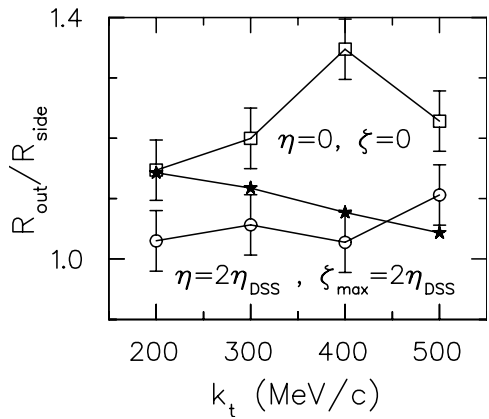}}
\caption{\label{fig:routrside}
Femtoscopic calculations of $R_{\rm out}/R_{\rm side}$ from a hybrid hydrodynamic/Boltzmann model for the case with viscosities and an initial anisotropy of the stress-energy tensor (circles) match expectations from STAR data (stars) \cite{Adams:2004yc}, while the ideal hydrodynamic calculation remains $\sim$15\% higher. Similar results were seen in \cite{romatschke}}
\end{figure}
To better understand these ideas, we have performed calculations based on the output of a radially- and boost-symmetric Israel-Stewart hydrodynamic code, which then feeds phase space points into a Boltzmann code that models the hadronic stage. Using color-glass initial conditions \cite{Hirano:2005xf,Drescher:2007cd} with an initial stress energy tensor set at $T_{xx}=T_{yy}=\epsilon/2$, and with Israel-Stewart relaxation times of 0.5 fm/$c$, calculations were performed with a simple equation of state that crudely mocks the soft region seen in lattice calculations for energy densities $\sim$ 1 GeV/fm$^3$ and the stiff region for higher energy densities where $c_s^2\sim 0.3$. Figure \ref{fig:routrside} compares two calculations, one for zero viscosity, and a viscous Israel-Stewart calculation where the initial anisotropy was chosen to be $T_{xx}=T_{yy}=\epsilon/2$, and $T_{zz}=0$. The anisotropies associated with the initial conditions and shear reduced the $R_{\rm out}/R_{\rm side}$ ratio by $\sim 15\%$.

Femtoscopic source radii are affected by a host of model choices and parameters, not just the viscosity. Among the features which have a non-neglible impact are:
\begin{itemize}
\item the initial profile -- color-glass initial conditions led to modestly lower $R_{\rm out}/R_{\rm side}$.\vspace*{-6pt}

\item the equation of state -- stiffer equations of state lower $R_{\rm out}/R_{\rm side}$ and decreases all three dimensions.\vspace*{-6pt}

\item adjusting the equation of state in the hadronic state to reflect non-equilibrium chemistry -- modestly inhibits surface emission and lowers $R_{\rm out}/R_{\rm side}$.\vspace*{-6pt}

\item relaxing the boost invariant assumption for longitudinal expansion -- this lowers $R_{\rm long}$ by 5-10\% \cite{Pratt:2008jj}.\vspace*{-6pt}

\item shear viscosity and early anisotropy of the field-dominated phase -- reduces $R_{\rm out}/R_{\rm side}$ by $\sim$20\%.\vspace*{-6pt}

\item bulk viscosity during the mixed phase -- increases all three radii.

\end{itemize}
The failure of some of the earlier hydrodynamic/Boltzmann models to reproduce $R_{\rm out}/R_{\rm side}$ does not derive from a single cause, but derives from numerous enhancements and model features, some of which change results by a few percent, while others change results by 10-20\%. 

It should be emphasized that determining bulk properties from femtoscopic analyses alone is naive. Many of the features and parameters from the list above also affect spectra and elliptic flow. For instance, the mean $p_t$ of protons was 10\% higher in the viscous calculation, as compared to the non-viscous calculations shown in Fig. \ref{fig:routrside}. Given the complex interplay between numerous observables and numerous features of the modeling, it is unlikely that rigorous quantitative conclusions about the bulk properties of matter can be extracted without a global model/data analysis. Nonetheless, the tremendous progress in modeling relativistic viscous hydrodynamics during the last year suggests that the model space is sufficiently flexible to find at least one point at which data can be reproduced. 
 
\section*{Acknowledgments}
Support was provided by the U.S. Department of Energy, Grant No. DE-FG02-03ER41259.

\vfill\eject
\end{document}